\documentclass[%
 reprint,
 amsmath,amssymb,
 aps,
pre,
]{revtex4-1}

\usepackage[pdftex]{graphicx}
\usepackage{dcolumn}
\usepackage{bm}
\usepackage{hyperref}

\usepackage{subfigure}
\usepackage{amsmath, amssymb}
\usepackage{enumitem}
\usepackage{color, colortbl}
\usepackage{afterpage}
\usepackage{url}
\usepackage{multirow}
\usepackage{rotating}
\usepackage{scalefnt}
\usepackage{hyperref}
\usepackage{ulem}
\usepackage{braket}
\usepackage{longtable}
\usepackage{xr}
\usepackage{soul}

\renewcommand{\(}{\left(}
\renewcommand{\)}{\right)}
\renewcommand{\[}{\left[}
\renewcommand{\]}{\right]}

\definecolor{asr}{rgb}{1., 1.,.2}
\definecolor{JB}{rgb}{1, 0.5, 0.5}
\definecolor{MM}{rgb}{0.5, 1, 0.5}
\definecolor{CT}{rgb}{0.4, 0.80, 1}

\begin{document}

\preprint{APS/123-QED}

\title{Revealing In-Block Nestedness: detection and benchmarking}

\author{Albert Sol\'e-Ribalta}
\affiliation{Internet Interdisciplinary Institute (IN3), Universitat Oberta de Catalunya, Barcelona, Catalonia, Spain}

\author{Claudio J.~Tessone}%
\affiliation{URPP Social Networks, Universit\"at Z\"urich, Switzerland}

\author{Manuel S.~Mariani}
\affiliation{Institute of Fundamental and Frontier Sciences, University of Electronic Science and Technology of China, Chengdu, PR China}
\affiliation{URPP Social Networks, Universit\"at Z\"urich, Switzerland}
\affiliation{Physics Department, Universit\'e de Fribourg, Switzerland}

\author{Javier Borge-Holthoefer}
\affiliation{Internet Interdisciplinary Institute (IN3), Universitat Oberta de Catalunya, Barcelona, Catalonia, Spain}
\affiliation{Institute for Biocomputation and Physics of Complex Systems (BIFI), Universidad de Zaragoza, Zaragoza, Spain}

\date{\today}

\begin{abstract}
As new instances of nested organization --beyond ecological networks-- are discovered, scholars are debating around the co-existence of two apparently incompatible macroscale architectures: nestedness and modularity. The discussion is far from being solved, mainly for two reasons. First, nestedness and modularity appear to emerge from two contradictory dynamics, cooperation and competition. Second, existing methods to assess the presence of nestedness and modularity are flawed when it comes to the evaluation of concurrently nested and modular structures. In this work, we tackle the latter problem, presenting the concept of \textit{in-block nestedness}, a structural property determining to what extent a network is composed of blocks whose internal connectivity exhibits nestedness. We then put forward a set of optimization methods that allow us to identify such organization successfully, both in synthetic and in a large number of real networks. These findings challenge our understanding of the topology of ecological and social systems, calling for new models to explain how such patterns emerge.

\end{abstract}

\pacs{
89.65.-s,	
89.75.Fb,	
89.75.Hc    
}
\maketitle


\section{\label{sec:level1}Introduction}

The identification of macroscale connectivity patterns has been central to the development of network science. 
Beyond the inherent methodological challenges of this task, ascertaining them is of relevance to the specific disciplines and to the area as a whole, inasmuch they are the outcome of distinct microscopic mechanisms of network formation. 
It is in this context --i.e.~understanding network architecture as an emergent feature-- that nestedness and modularity arise as prominent macro-structural signatures to study.

\begin{figure}[t]
  \centering
    \includegraphics[width=\columnwidth]{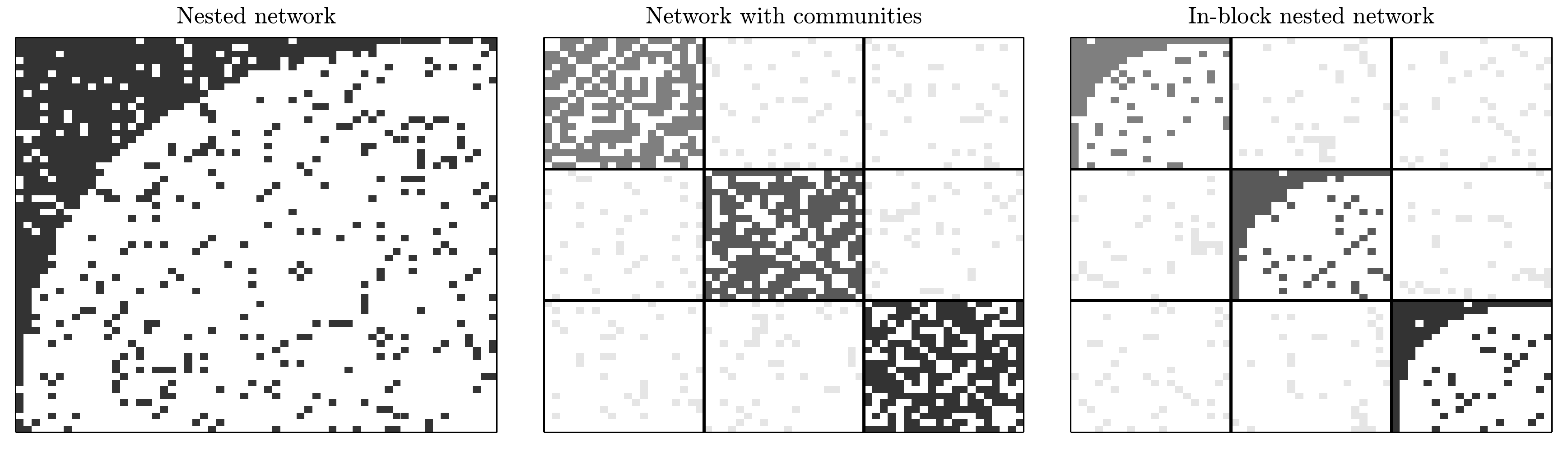}
  \caption{{\bf Left:} Example of network with \textit{nested structure}. Rows and columns have been ordered by degree. {\bf Center:} Example of a network with \textit{block structure} -- \textit{blocks} are typically referred to as \textit{communities} and \textit{compartments} in the network science~\cite{fortunato2010community} and the ecology~\cite{prado2004compartments} literature, respectively. The network exhibits high internal connectivity between nodes of the same block and low connectivity between nodes that belong to different blocks. {\bf Right:} Example of a network with \textit{in-block nested (IBN) structure}. Nodes within blocks exhibit nested structure. Vertical and horizontal lines are a visual guide to show the existing blocks.}
   \label{fig:network_examples}
\end{figure}

The concept of \textit{nestedness} was first coined in biology to characterize the spatial distribution of biotas in 
isolated, yet spatially-related landscapes~~\cite{patterson1986patterson}, and later found to describe large families of inter-species cooperative relations~\cite{bascompte2003nested}. 
In structural terms, a perfectly nested pattern is such that the set of connections of any given node is a  subset of the relationships of larger degree ones~\cite{bipartite}; see Fig.~\ref{fig:network_examples} (left). 
Nestedness has imposed itself as a landmark feature in mutualistic interactions, with an emphasis in natural ecosystems, triggering a large amount of research spanning fieldwork~\cite{bascompte2003nested}, modeling~\cite{saavedra2009simple} and simulation~\cite{bastolla2009architecture}. 
Beyond natural systems, nestedness emerges as well in social, technical, and economic systems, e.g.~industrial relationships~\cite{Uzzi1996b,saavedra2009simple,bustos2012dynamics}, international trade~\cite{Saracco2015}, information ecosystems~\cite{borge2017emergence}, anthropology~\cite{kamilar2014cultural} and knowledge production ~\cite{Cimini2014}.
In socio-economic systems, epitome of this property in unipartite networks, emergence of nestedness is originated in agents attempting to maximize their own centrality ~\cite{koenig11,Bardoscia2013,konig2014nestedness}.

On the other side, the identification of \textit{modular} patterns in networks stands as one of the hallmarks 
in the area with prominent precedents in social network analysis~\cite{coleman1964introduction}. Besides social systems, networks with significant community structure, see Fig.~\ref{fig:network_examples} (middle), appear in multiple contexts~\cite{fortunato2010community}, like  biology~\cite{guimera05} or cognitive science~\cite{borge2010semantic}. It implies the existence of subgroups of nodes, strongly connected within but loosely connected to nodes outside. 
The identification and analysis of community structure constitutes itself a sub-area of network science. It poses challenges with respect to detection algorithms, empirical problems, applications and conclusions derived~\cite{Fortunato2016}.

Nestedness and modularity have been often treated as incompatible architectures, since they are thought to emerge from conflicting (respectively, cooperative and competitive) dynamics~\cite{thebault2010stability}. Thus, most studies have focused exclusively on either of them. The existence of systems which combine both patterns has been largely overlooked, despite challenging indications in natural~\cite{prado2004compartments,lewinsohn2006structure,fortuna2010nestedness,flores2013multi} and social ecosystems~\cite{borge2017emergence}. As of now, the proper identification of such compound structures lays beyond the capabilities of state-of-the-art techniques. 

In the scarce existing literature we identify two different approaches. 
The first  operates in parallel, measuring modularity and nestedness independently ~\cite{fortuna2010nestedness,olesen2007modularity,borge2017emergence}, with the obvious drawback that these properties are treated as emerging unrelatedly. 
The second approach operates sequentially: After a proper identification of a partition (usually in terms of modularity~\cite{newman04a}), it computes the nestedness (usually in terms of NODF~\cite{almeida2008consistent}) locally for each block ~\cite{flores2013multi,lewinsohn2006structure}. In consequence, modularity takes functional precedence relegating nestedness from a macro- to a mesoscopic pattern. Both approaches overlook that these two network patterns are inherently intertwined and thus cannot be evaluated using \textit{independent} metrics. 

\begin{figure*}[ht]
	\begin{center}
		\includegraphics[width=2\columnwidth]{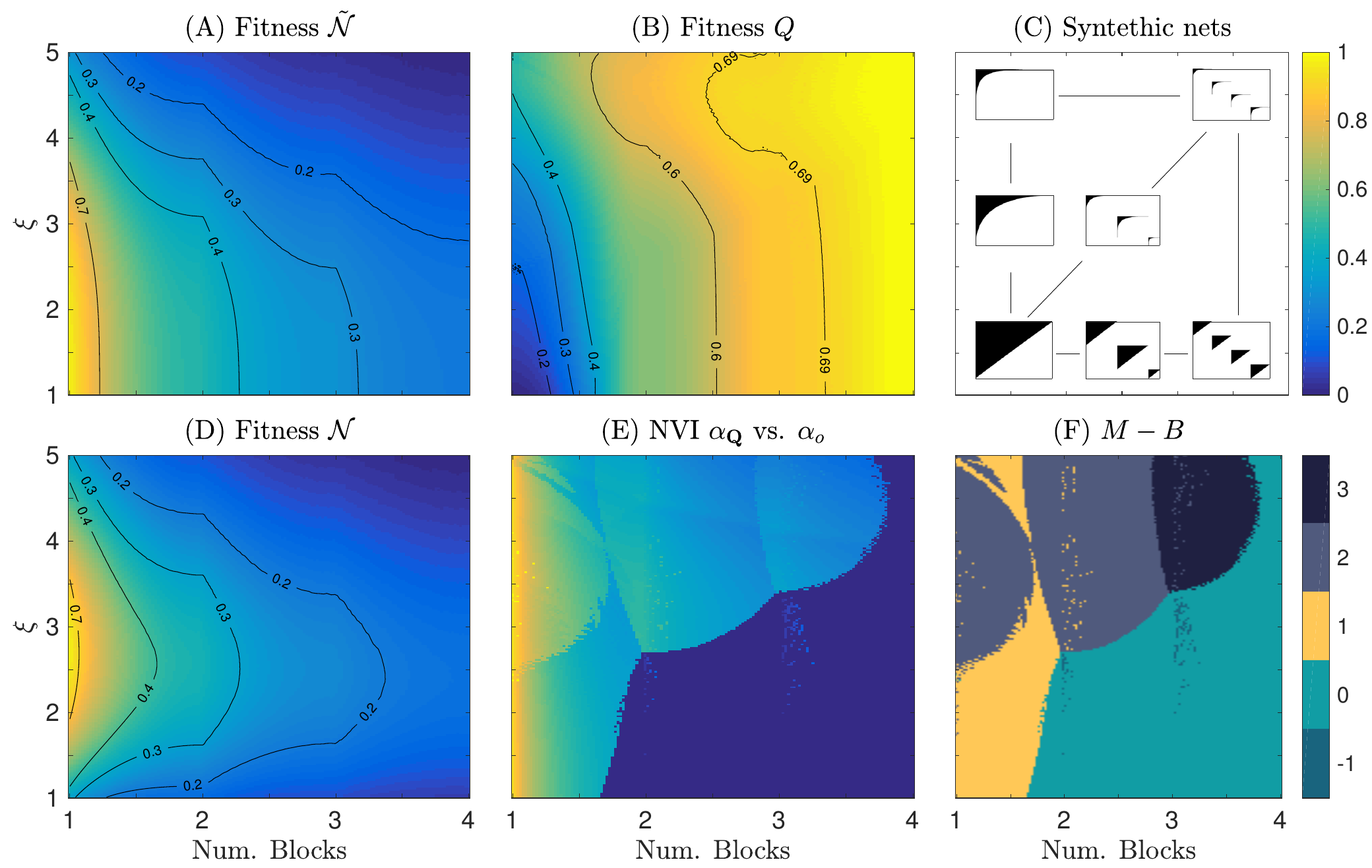}
	\end{center}
	\caption{Behavior of nestedness metrics $\widetilde{\mathcal{N}}$ and $\mathcal{N}$, and modularity $Q$ in noiseless IBN synthetic networks. Panel (A) illustrates how global nestedness $\widetilde{\mathcal{N}}$ decreases with the number of blocks in a network --regardless of their internal perfectly nested structure. Notice that the shape parameter $\xi$ also affects $\widetilde{\mathcal{N}}$, as a side-effect of the strict decreasing condition imposed in the original definition of NODF. Panel (B) illustrates how modularity $Q$ increases with the number of blocks in a network. Panel (C) illustrates the kind of networks generated by the model (see Section \ref{nemo_dataset}). The vertical axis represents the shape parameter $\xi$, that quantifies the slimness of the nested structures. The horizontal axis stands for the number of modules. In non-integer settings (e.g. 2.5) there are an integer amount of nested blocks of the same size (e.g. 2) and one block of relative size equal to the remaining fraction (e.g. 0.5 of the size of the other communities). Panel (D) shows the global nestedness fitness $\mathcal{N}$ which incorporates a null model, as defined by Eq.~\ref{Nhat}. Panel (E) shows the normalized variation of information between the modules detected by modularity optimization and the ground- truth (GT) blocks. Panel (F) shows the difference $M-B$ between the number $M$ of modules detected by the modularity optimization and the number $B$ of ground-truth blocks.}
	\label{fig:mod_nest_indep}
\end{figure*}

To be precise, the presence of modules places hard limits to the extent of nestedness that a network can exhibit (Fig.~\ref{fig:mod_nest_indep}A); and, on the other direction, detected communities in a globally nested system leads to aberrant, hardly interpretable modules ~\cite{lewinsohn2006structure} (see Fig.~\ref{fig:mod_nest_indep}F and the relative discussion in the Results section). As it is expected, the modularity score is sensitive to the number of communities (Fig.~\ref{fig:mod_nest_indep}B), but not to the shape of the nested structure. Overall, these {\it in-block nested} (IBN) structures are highly undetectable if the network contains few communities and/or the nested structure within the communities is very stylized (Fig.~\ref{fig:network_examples}, right). 

Beyond the methodological challenges, there exists an important epistemological aspect which cannot be overlooked.
In most scenarios, the boundaries of the system under consideration are imprecise because the researchers, albeit involuntarily, impose a discretionary observation scale to it. 
Extending the realm of observation, the network structure can parsimoniously be expected to show a set of loosely interconnected  blocks. 
This is particularly evident in natural ecosystems~\cite{flores2011statistical}, where, in general, no precise geographic boundaries can be defined; but also in social networks, when it comes to decide which subjects should be included or not in a specific study.

The paper is organized as follows: to overcome the limitations of existing approaches, Section \ref{sec:defs} introduces a compact methodological framework that jointly considers both patterns (block --or compartmental-- structure and within-block nestedness). Our methodology can unveil the existence of IBN structures, as shown in Section \ref{sec:synth_nets} for a suitable benchmark. We then investigate, in Section \ref{sec:real}, the question of how general (or anecdotal) such property is. We show that a large number of real datasets exhibit in-block nested structures that would have gone undetected under conventional modularity optimization/nestedness detection techniques. Our findings indicate that these previously-overlooked structures are in fact common in ecological and social systems. This opens a new direction for the structural analysis of ecological and social systems, discussed in Section \ref{sec:conclusion}, calling as well for new models to explain how IBN structures emerge.

\section{Definition and quantification of in-block nestedness}
\label{sec:defs}
In this Section, we develop a proper formulation of the problem of determining to what extent a given network is organized as loosely interconnected \textit{blocks}, each of them internally nested.
We begin by defining in a congruent manner both, global nestedness and the new in-block nestedness fitness $\mathcal{I}$. 
In particular, maximizing the IBN fitness function $\mathcal{I}$ allows us to unveil the best node partition in terms of IBN structure.
We analyze synthetic networks to show that $\mathcal{I}$-maximization allows us to reconstruct ground truth IBN structures that would have gone undetected under the widely-used modularity optimization.
We then proceed to analyze a large set of real networks -- originated in the most varied disciplines -- to evince that this type of structures is indeed a common occurrence in both uni- and bipartite networks  of diverse nature.

\subsection{A nestedness measure}
Without loss of generality, consider a bipartite network, describing a relationship between two sets $G = \{s,t,\dots \}$ and $\Gamma = \{\sigma,\tau,\dots \}$ with cardinalities $N_r$ and $N_c$ respectively. The bipartite network can be represented as a binary adjacency matrix $\mathbf{A}$ whose elements are $A_{s,\tau} = 1$ if a relationship between elements $s$ and $\tau$ exists, or zero otherwise.
In the same spirit as the measure Node Overlap-Decreasing Fill~\cite{almeida2008consistent} (NODF), we introduce the global nestedness fitness $\widetilde{\mathcal{N}}$ which measures the amount of global overlap between row and column pairs
\begin{eqnarray} \label{eq:NODF}
	\widetilde{\mathcal{N}} &=& \frac{2}{N_r+N_c} \left\{ \sum_{s,t}^{N_r} \frac{O_{s,t}}{k_t \(N_r-1\)} \Theta(k_s-k_t)  \right. \nonumber \\
	 & &+ \left. \sum_{\sigma,\tau}^{N_c} \frac{O_{\sigma,\tau}}{k_\tau \(N_c-1\)} \Theta(k_\sigma-k_\tau) \right\},
\end{eqnarray}
where $k_i$ corresponds to the degree of the element $i$ (regardless on whether it belongs to $G$ or $\Gamma$);
$\Theta( \cdot )$ is the Heaviside step function (such that the only contributing terms  are those in which the outer index has larger degree than the inner); $O_{\cdot,\cdot}$ measures the degree of overlap between row and column pairs as:
\begin{eqnarray} \label{eq:NODF_OV}
	O_{s,t} = \sum_{\upsilon=1}^{N_r} A_{s\upsilon}A_{t\upsilon}, \;\;\;\;\;\;
	O_{\sigma,\tau} = \sum_{u=1}^{N_c} A_{u \sigma }A_{u \tau}. 
\end{eqnarray}
It is important to remark that Eq.~\ref{eq:NODF} weighs linearly the contribution of rows and columns to $\widetilde{\mathcal{N}}$ (instead of quadratic weights as in NODF~\cite{almeida2008consistent}).
This is preferable when the difference between the number of rows and columns is considerable.

\subsection{Recasting nestedness at the mesoscale level}

We now introduce two new elements that allow us to generalize $\widetilde{\mathcal{N}} $ to the case where nested structures exist at a mesoscopic scale: a membership variable and a null model. 
First, we consider that both sets of nodes are partitioned into $C$ disjoint subsets, termed \textit{blocks}.
This implies that for each node $i$, it is possible to define a membership variable $\alpha_i$.
Based on this, the total size of block $\ell$ can be obtained as
\begin{eqnarray}
	C(\ell) = \sum_{s=1}^{N_r} \delta(\alpha_s, \ell) + \sum_{\sigma=1}^{N_c} \delta(\alpha_\sigma, \ell), 
\end{eqnarray}
where $\delta$ is the Kronecker Delta. In addition, $C_s = \sum_t \delta(\alpha_s, \alpha_t) $ and  $C_\sigma = \sum_\tau \delta(\alpha_\sigma, \alpha_\tau)$ give, respectively, the number of nodes in the block nodes $s$ and $\sigma$ belong to. 
The block overlap $O_{\cdot,\cdot}$, now including the membership variable, can be obtained as 
\begin{eqnarray*} \label{eq:NeMo_ov}
	O_{s,t} & = &\sum_{\upsilon=1}^{{N_c}} A_{s\upsilon}A_{t\upsilon}\delta(\alpha_s,\alpha_\upsilon),\\
	O_{\sigma,\tau} &= &\sum_{u=1}^{{N_r}} A_{u\sigma}A_{u\tau}\delta(\alpha_u,\alpha_\sigma). 
\end{eqnarray*}
 
The null model gauges the expected overlap between a pair of nodes belonging to a class and aims at compensating the nestedness that can be explained solely by the nodes' degrees. The expected overlap, given two nodes $s$ and $t$ with degrees $k_s$ and $k_t$, is obtained considering that the neighbors of each node are chosen uniformly at random. In this situation, the probability that both nodes share a common neighbor is simply $({k_{s}}{k_{t}})/{{N_c}^2}$ and therefore, the expected amount of shared neighbors, i.e.~the expected overlap, is given by
$\langle O_{s,t} \rangle = {k_{s}k_{t}}/{{N_c}}$.
The same argument shows that for the columns,  $\langle O_{\sigma,\tau} \rangle = k_\sigma k_\tau / {N_r}$.

We can now introduce the in-block nestedness fitness $\mathcal{I}$, which quantifies to which extent a network exhibits IBN, 
\begin{widetext}
\begin{eqnarray} \label{eq:NeMo}
	\mathcal{I} = \frac{2}{{N_r}+{N_c}} \left\{ \sum_{s,t}^{{N_r}}\[\frac{O_{s,t} - \langle O_{s,t} \rangle}{k_t \(C_s-1\)} \Theta(k_s-k_t)\delta\(\alpha_s,\alpha_t\)\]  	
	 + \sum_{\sigma,\tau}^{{N_c}}\[\frac{O_{\sigma,\tau} - \langle O_{\sigma,\tau} \rangle}{k_\tau  \(C_\sigma-1\)}\Theta(k_\sigma-k_\tau)\delta\(\alpha_\sigma,\alpha_\tau\)\] \right\} .
\end{eqnarray}
\end{widetext}
In this expression, some normalization factors have disappeared after straightforward simplifications.
In the same spirit as in the row and column weighting of $\widetilde{\mathcal{N}}$ (Eq.~\ref{eq:NODF}), the per-block  nestedness aggregates are  weighted by the size of the block (i.e.~$C_s$ and  $C_\sigma$). Notice that, for each pair of row nodes, $O_{s,t}$ only accounts for column nodes within the same block, while $\langle O_{s,t} \rangle$ considers all column nodes regardless of the block they belong to. This implies that, for any pair of rows, $\mathcal{I}$ will be in principle larger when they are assigned to the same block: in this case the difference $O_{s,t}-\langle O_{s,t} \rangle$ has positive contributions. On the other hand, the membership variable $\alpha$ allows to discard some of the comparisons, assigning row nodes to different communities. In general, an algorithm that correctly  maximizes $\mathcal{I}$ will attempt to discard pairs whose contribution is negative to the aggregate. This intuition is equivalent for columns. The balance of such ``merge-split'' strategy for rows and columns allows an algorithm to identify in-block nested structures by maximizing the objective function $\mathcal{I}$. Equation \ref{eq:NeMo} is equally valid for unipartite networks, simply imposing that the two sets of nodes are identical. In this work, we have adopted a biologically-inspired optimization algorithm~\cite{karaboga2005idea}. However, $\mathcal{I}$'s formulation --which closely follows that of modularity $Q$-- enables the adoption of many existing heuristics (see \cite{fortunato2010community} for an extensive review).

Noteworthy, the objective function $\mathcal{I}$ reduces to $\widetilde{\mathcal{N}}$, corrected by a suitable null model, if one considers a single block ($\alpha_{s} = \alpha_{\sigma} = \alpha, \forall s,\sigma$), i.e.
\begin{eqnarray*} \label{Nhat}
\mathcal{N} &=& \frac{2}{{N_r}+{N_c}} \left\{ \sum_{s,t}^{{N_r}} \frac{O_{s,t} - \langle O_{s,t} \rangle}{k_t (N-1)} \Theta(k_s-k_t)  \right. \\	
	& & + \left.  \sum_{\sigma,\tau}^{{N_c}} \frac{O_{\sigma,\tau} - \langle O_{\sigma,\tau} \rangle}{k_\tau  ( M-1 )}\Theta(k_\sigma-k_\tau)  \right\}.
\end{eqnarray*}
While the difference between NODF and $\widetilde{\mathcal{N}}$ is slight --except when $M \ll N$, or viceversa--, the null-model correction in $\mathcal{N}$ heavily alters the nestedness measure. In particular, note that fully connected nodes do not contribute to $\mathcal{N}$ --as opposed to maximum contribution in the original formulation. 

To illustrate this point, it is instructive to consider the case of one single nested block (Num. Blocks = 1 in Fig.~\ref{fig:mod_nest_indep}A,C,D) (Note that Fig.~\ref{fig:mod_nest_indep} is explained in detail in the next Section). In this cut, the contrast between non-corrected measure of nestedness (panel \ref{fig:mod_nest_indep}A) and corrected (panel \ref{fig:mod_nest_indep}D) is very clear: when the perfectly nested network is very dense (bottom-left corner, $\xi < 2.5$), most of the nodes have large expected overlap with the few hubs. Hence, even though the nestedness condition is respected for all the pairs of nodes, the nestedness metric $\widetilde{\mathcal{N}}$ only deviates little from its expected value under the null model, resulting in a small value of $\mathcal{N}$.
In other words, the observed level of nestedness $\widetilde{\mathcal{N}}$ can be simply explained by the network degree distribution.
On the contrary, $\widetilde{\mathcal{N}}$ and $\mathcal{N}$ are practically identical for $\xi > 2.5$. In this region, the slimness of the nested structure and the strict decreasing connectivity condition (Heaviside function in Eqs.~\ref{eq:NODF} and \ref{eq:NeMo}) heavily limit the value of nestedness, regardless the consideration of a null model ($\mathcal{N}$) or not ($\widetilde{\mathcal{N}}$).

\section{Detection of in-block nested structures in synthetic networks} \label{sec:synth_nets}
\label{sec:synth_nets}

In this Section, we first introduce a benchmark graph model with planted in-block nested structures (Section \ref{nemo_dataset}), and then present the results for the $Q$ and $\mathcal{I}$ optimization algorithms' performance in reconstructing such planted structures (Section \ref{sec:ibn_opt}).

\subsection{In-block nested structures generator} \label{nemo_dataset}

In a perfectly nested structure, rows (and columns) interact with a subset of the neighbors of the rows (and columns) of larger degree. Correctly ordering its adjacency matrix by row and column degree, it resembles an upper (possibly with some curvature) secondary diagonal matrix. Inspired by the $p$-norm unit ball equation, we synthetically generate such structures using 
\begin{eqnarray}\label{nestedShapeFunction}
	y=f_{n}(x)=1-(1-x^{1/\xi})^\xi\,
\end{eqnarray}
where $x \in [0,1]$ and $\xi \in [1,\infty)$ dictates how stylized is the shape of the nested structure. The adjacency matrix of the nested structure with $N_c$ nodes is constructed tessellating the $[0,1] \times [0,1] $ space into $N_c^2$ squares; and then, adding a link into each matrix position whose center lies above the curve in Eq.~\ref{nestedShapeFunction}. 
While an approach based on threshold graphs would have been also used, the degree sequences it produces are  stochastic, introducing unnecessary fluctuations to the network generation \cite{Grimm2017}. 

Such noiseless nested structures are rarely found in real systems. Thus, on top of the previous scheme, we mimic random and uncorrelated noise using a dual-step procedure. In the first stage,  we randomly remove links from the perfectly nested structure with probability $p$. Given a network with $E_c$ edges, $pE_c$ of them will be removed in average. In the second step, the removed edges are randomly distributed across the empty elements of the adjacency matrix. These include initially empty positions (i.e.~those lying below the function in Eq.~\ref{nestedShapeFunction}) and empty positions resulting form the stochastic removal procedure. Parameter value $p=0$ corresponds to the initial structure and $p=1$ corresponds to an Erd{\H{o}}s-R\'{e}nyi network with average degree $E_c/N_c$.

The construction of an adjacency matrix of an ideal in-block nested structure starts off with $B$ (a real-valued number) blocks. Specifically, we build $\lfloor B \rfloor$ blocks of size $\lfloor N_c/ B \rfloor$ and another with the remaining  $N_c - \lfloor N_c/ B \rfloor$ nodes. In the previous, $\lfloor \cdot \rfloor$ stands for the integer part function. In this way, the network produced has some level of heterogeneity (albeit the size of all blocks remains in the same order of magnitude).
Then $A_{\mathcal{I}}$, can be constructed repeating the procedure for each block, $A^{c}$, and joining them to compose a block diagonal matrix
\begin{eqnarray}\label{inblockneststructure}
A_\mathcal{I} = 
	\begin{bmatrix}
		A^{c_1}&A^{o_{11}}&\cdots & A^{o_{1B}}\\
		A^{o_{21}} &A^{c_2} &\cdots & A^{o_{2B}}\\
		\vdots & \vdots & \ddots & \vdots\\
		A^{o_{B1}} & A^{o_{B2}} & \cdots & A^{c_B}\\
	\end{bmatrix}\,,
\end{eqnarray}
where $A^{o} = {\bf 0}$ is a matrix of the required size.

Similarly to the intra-block noise, we reproduce inter-block perturbations with an additional dual-step procedure controlled by the parameter $\mu \in [0, 1]$. In this first step, for each block, each link is removed with probability $p_i = \mu(B-1)/B$. In the second step, those links are distributed at random to connect one node of the original block with a random node of a different block. Probability $p_i$ depends on the number of blocks, since our purpose is that for $\mu=1$ the amount of links within each block is the same as the amount of links connecting any two distinct blocks. In the limit situation of $p=\mu=0$ the outcome corresponds to a noiseless in-block nested structure and for $p=\mu=1$ the outcome corresponds to an Erd{\H{o}}s-R\'{e}nyi network with the same average degree as the germinal noiseless structure, see Appendix \ref{sec:syntavdeg}.

The described generative process can be locally implemented in terms of edge probabilities. Along these lines, the probability of having a link between nodes $i$ and $j$ within a block becomes
\begin{eqnarray}\label{intracomm_link}
	P(A^c_{ij})= && \left[(1-p + p\,p_{r})\, \Theta(j\, N_c - f_n(i\, N_c)) \right. +  \\ 
	&& \left. p_{r} \,(1-\Theta(j\, N_c - f_n(i\, N_c)) \right] \(1-p_i\),\nonumber 
\end{eqnarray}
where $\Theta$ is the Heaviside function. The term within square brackets is related to the intra-block noise. In the first term, $(1-p)$ corresponds to the probability of not altering the link. The second, $pp_{r}$, corresponds to the probability of recovering a link, after removal, in the random dispersion of removed links. These two terms are restricted, by $\Theta$ function, to the region where links exist in the noiseless structure. The third term, $p_r=pE_c(N_c-E_c+pE_c)^{-1}$, corresponds to the probability of selecting link $A_{ij}$ in the random distribution of removed links. Eventually, the term $(1-p_i)$ corresponds to the probability of not removing the link in the process of generating inter-block noise.

The probability of a inter-block link is
\begin{eqnarray}\label{intercomm_link}
	P(A^o_{ij})&&=\frac{2E_c p_i}{2(B-1)N_c^2}=\frac{\mu E_c}{N_c^2B}.
\end{eqnarray}
The numerator accounts for the amount of removed links from the blocks related to the off-diagonal block $A^o$, that is compartment $i$ and $j$. The denominator accounts for the possible places where each of those links can be placed. Note that the $2$ on both numerator and denominator explicitly shows that each removed link of compartment $A^{c_k}$ can be reallocated in $A^{o_{k\cdot}}$ or $A^{o_{\cdot k}}$.

The noisy version given by Eqs. \ref{intracomm_link} and \ref{intercomm_link} of the original noiseless in-block nested structure generates a network with equivalent average degree. This is formally proved in Appendix \ref{sec:syntavdeg}. An example of these synthetically generated structures is shown in Fig.~\ref{fig:mod_nest_indep}C and Fig.~\ref{fig:nestmod_comparison}C.

\subsection{IBN optimization applied to synthetic networks}
\label{sec:ibn_opt}
\begin{figure*}[bt]
	\begin{center}
		\includegraphics[width=2\columnwidth]{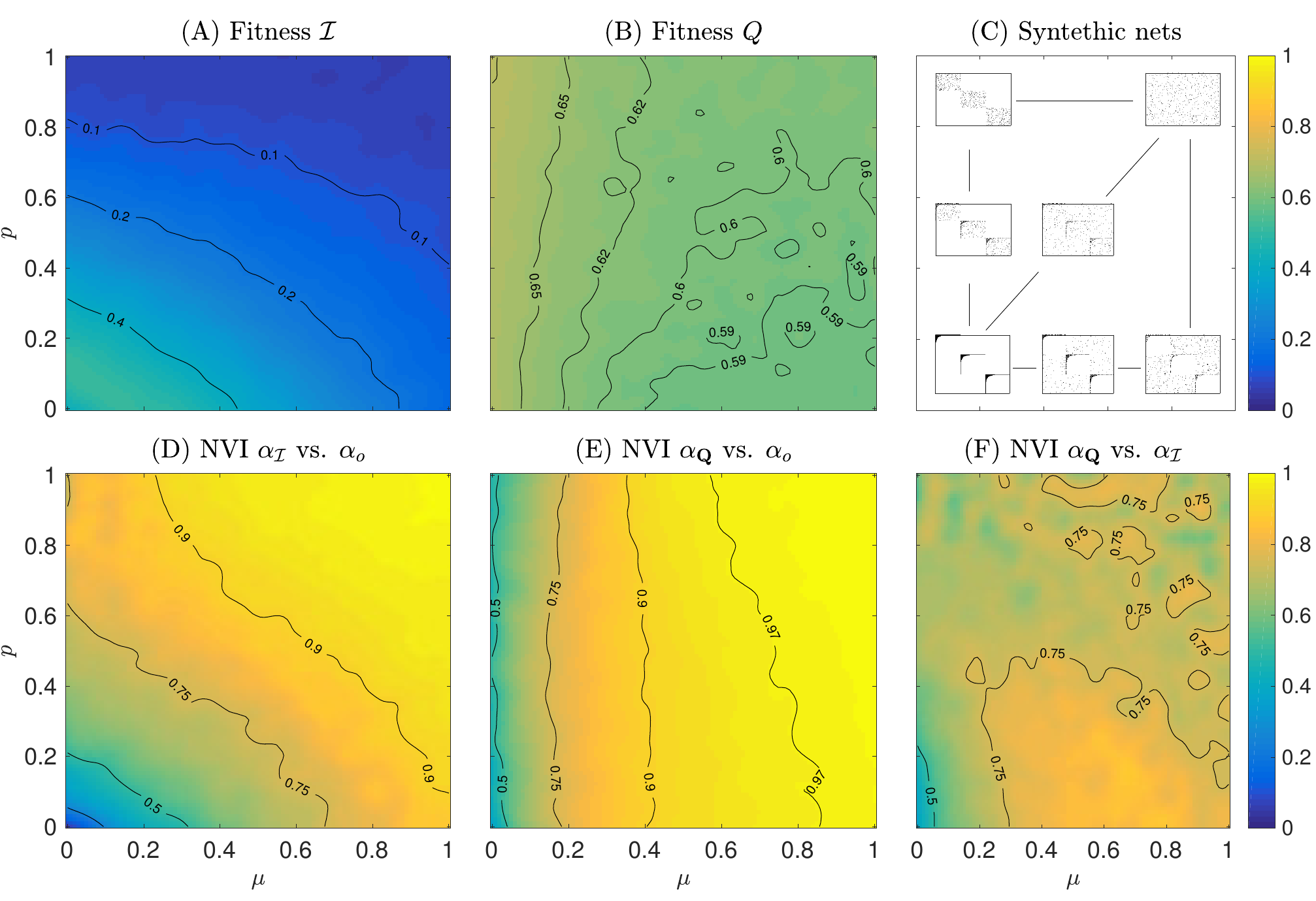}
	\end{center}
	\caption{Results for synthetic IBN networks with $B=3$ and $\xi=4$, for varying values of $p$ and $\mu$. Panel (A) displays the value of the fitness function $\mathcal{I}$ as obtained with a biologically-inspired optimization algorithm. Similarly, panel (B) reports the value of modularity $Q$ as obtained with the Combo optimization algorithm ~\cite{sobolevsky2014general}. Panels (D) and (E) compare the $\mathcal{I}$-optimized ($\alpha_{\mathcal{I}}$) and the $Q$-optimized ($\alpha_Q$) partitions, respectively, to the planted partition ($\alpha_0$) via the NVI. Panel (C) provides a visual intuition of the effect of parameters $p$ and $\mu$ on the synthetically-generated networks. 
    Panel (F) shows the difference between the $\mathcal{I}$-optimized ($\alpha_{\mathcal{I}}$) and the $Q$-optimized ($\alpha_Q$) partitions as measured by their NVI.}
	\label{fig:nestmod_comparison}
\end{figure*}

In Fig.~\ref{fig:mod_nest_indep}, we unveil the limitations of current techniques to detect IBN structures in noiseless networks (IBN networks) where no links exist between nodes  belonging to different blocks. In a more realistic setting, where 
links can connect nodes that belong to different blocks, such weaknesses become even more apparent (see Fig. \ref{fig:nestmod_comparison}). 

We generated synthetic (unipartite) networks of $N_c$ nodes and $E_c$ edges within blocks, where the level of in-block nestedness and the number of inter-block links can be varied in controlled manner by means of few parameters (see Methods): the number of blocks $B$, the shape parameter $\xi$, the in-block nestedness parameter $p$, and the mixing parameter $\mu$.
To allow for heterogeneity in block size, we build $\lfloor B \rfloor$ blocks of size $\lfloor N_c/ B \rfloor$ and another with the remaining nodes. $\xi$ determines the density of the network, controlling how stylized the nested structure is.
The in-block nestedness parameter $p$ gives the fraction of links that do not respect the notion of perfectly nested organization within a block and 
the mixing parameter $\mu$ measures the fraction of inter-block links.

The experiments in Fig.~\ref{fig:mod_nest_indep} correspond to $3 \times 10^{4}$ networks generated with parameters $p = \mu = 0$, with varying number of blocks and number of edges (see Fig.~\ref{fig:mod_nest_indep}C for an illustration of the resulting adjacency matrices). By construction all networks have maximum ${\mathcal I}$, in the range (0.16, 0.90), depending on the shape parameter $\xi$ and number of blocks $B$.
Unsurprisingly, modularity $Q$ increases as the number of block increases (Fig.~\ref{fig:mod_nest_indep}B). We find that in this setting, the modules detected by the modularity-maximization algorithm can be very different from the planted blocks, as measured by Normalized Variation of Information~\cite{meila2003comparing} (NVI), Fig.~\ref{fig:mod_nest_indep}E. The difference between detected and planted blocks is larger for sparser networks (upper region of Fig.~\ref{fig:mod_nest_indep}E).
Fig.~\ref{fig:mod_nest_indep}F shows the difference between the number $M$ of detected modules by the $Q$-maximization and the number $B$ of planted blocks. The difference is non-zero for a large region of the parameter space.
In particular, the modularity optimization algorithm detects more than one module in a network composed of one single block with internal nested structure (see Fig.~\ref{fig:mod_nest_indep}F, left-corner).
This happens because the modularity-optimization algorithm tends to form a module that only contains the nodes with largest degree. Figs.~\ref{fig:mod_nest_indep}E-F indicate that modularity optimization is only reliable in the limit of large number of blocks and dense networks (lower-right corner).

The results of Figs.~\ref{fig:mod_nest_indep}B,E,F make clear that measuring modularity and nestedness as two independent network properties is inherently flawed: modularity-optimization algorithms detect more than one module in a network composed of a single nested block, and the NVI between detected modules and planted blocks is in general large.
We have verified that the $\mathcal{I}$-optimization algorithm introduced in this paper overcomes these limitations and is able to correctly recover the planted structure for all the parameter values shown in Fig.~\ref{fig:mod_nest_indep}.

Figure~\ref{fig:nestmod_comparison} shows the results for an exhaustive exploration of the $(p, \mu)$ parameter space over 2600 networks with a fixed $B=3$ and shape parameter $\xi=4$ (see Fig.~\ref{fig:nestmod_comparison}C for an illustration of the resulting adjacency matrices).
Results on these synthetic networks after a modularity optimization process (Fig.~\ref{fig:nestmod_comparison}B) show that $Q$ is almost insensitive to changes in the parameters of the model: notice that its range is quite narrow, $0.55 < Q < 0.7$, and only mildly affected by the $p$ parameter (i.e., by the level of IBN). This itself is a consequence of the fact that $Q$ does not consider any particular structure within the blocks, but only their internal density. Even further, the value of modularity fluctuates around $Q \approx 0.6$, in remarkable accordance with the predictions in Guimer\`a {\it et al}.~\cite{guimera2004modularity}, for sparse graphs, like those obtained with $\xi=4$. 

Figure~\ref{fig:nestmod_comparison}A shows the value of in-block nestedness fitness, $\mathcal{I}$, after a maximization procedure based on a biologically-inspired optimization algorithm (see Methods). Evidently, $\mathcal{I}$ is sensitive to both the modular structure, and the nested organization within, taking a maximum value for $\mu=0$ and $p=0$. When we increase the randomness in either dimension the obtained in-block nestedness fitness smoothly decreases, reaching a global minimum when $\mu = 1$ and $p = 1$. 

Since it can be assumed that the block structure is known {\it a priori}, it is possible to quantify how far a given partition is from the planted one. As before, we resort on NVI to assess the quality of the partitions obtained optimizing $\mathcal{I}$ (Eq.~\ref{eq:NeMo}) and the quality of the partitions obtained maximizing $Q$~\cite{sobolevsky2014general}. 
Focusing on the quality of the $Q$-detected partition with respect to the prescribed one (Fig.~\ref{fig:nestmod_comparison}E), we emphasize that modularity does not recover the planted partition in any parameter configuration, not even at $\mu=0$. Remarkably, changes in NVI are independent of the parameter $p$, related to the level of disorder within each block.
In contrast, we see that $\mathcal{I}$ optimization allows to unveil the planted partition for a region along the $\mu$ axis, as long as $p$ remains low (Fig.~\ref{fig:nestmod_comparison}D): the presence of internal nestedness compensates the tenuous identity of the blocks, caused by large $\mu$. The parameter region corresponding to low $p$ and large $\mu$ is also the region where $Q$-detected partitions and $\mathcal{I}$-detected partitions differ the most (Fig. \ref{fig:nestmod_comparison}F). This points out that the $Q$-detected partitions are particularly unreliable when there is a clear internal nested structure and there exist a significant number of inter-block links.


\section{Detection of in-block nested structures in real datasets}
\label{sec:real}
The previous sections demonstrate the adequacy and robustness of ${\mathcal I}$ --and the inherent flaws of modularity $Q$ 
and global nestedness $\mathcal{N}$-- to unveil IBN structures.
However, those analysis would be limited to a mere academic exercise in absence of ample (in terms of examples and origin) empirical evidence. 
To demonstrate the practical aspects of the proposed methodology, we have analyzed a total of 334 networks, including both unipartite (57) and bipartite (277) ones which are known to display some level of nested organization. Most of them ($209$ bipartite networks) belong to ecology~\cite{weboflife} --mostly mutualistic networks-- and the rest belong to online platforms ($68$ bipartite networks) and social networks ($57$ unipartite networks). Table \ref{tab:netdescr} in Appendix \ref{app2} details the origin and characteristics of each dataset.

\begin{figure}[t]
	\begin{center}
		\includegraphics[width=1\columnwidth]{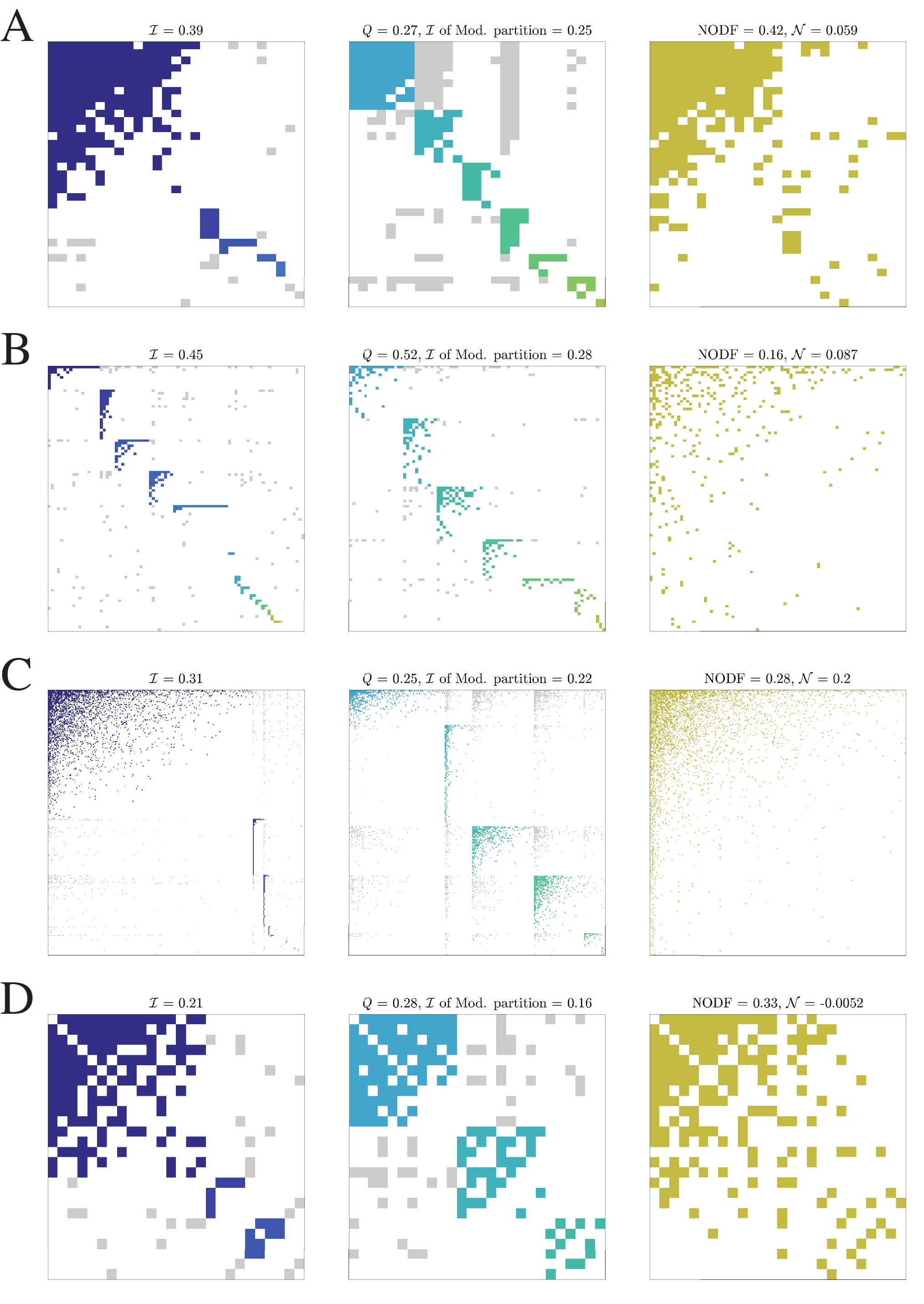}
	\end{center}
	\caption{Interaction matrices of four illustrative systems: (A) a host-parasite competitive bipartite network in Volga-Kama Nature Reserve; (B) a pollination mutualistic bipartite network in Cord\'on del Cepo, Chile; (C) a urban bipartite network accounting for citizen visits to city services in Chennai, India; and (D) a unipartite friendship network in a Dutch school class. For all of them, rows and columns have been arranged to highlight different aspects: block membership and degree for $\mathcal{I}$- (left) and $Q$-maximization (center) partitions; global degree ranking in the right column.}
	\label{fig:real_results2}
\end{figure}

As a visual intuition, Fig.~\ref{fig:real_results2} displays the adjacency matrix of four of these networks, where rows and columns have been sorted following different criteria: for left and central columns ($\mathcal{I}$- and $Q$-maximizing partitions, respectively), nodes in the same block are placed together, and they are ranked by degree (within blocks) to make more apparent a possible IBN structure; in the right column, nodes are simply ranked by degree. 
Panel A shows such arrangements for a host-parasite network (see A\_HP\_050 in Table \ref{tab:netdescr} of the Appendix \ref{app2}). Clearly, the three matrix representations look very different. In this case, $\mathcal{I}$ favors the existence of a large, highly nested block, and a set of smaller clusters with a clear internal organization as well, whereas $Q$ renders several, similarly sized, highly dense modules with no clear internal nested organization. Even though the classical NODF measure hints at some degree of global nestedness, taking into account the null model ($\mathcal{N} = 0.059$) seems to indicate that the nested organization is a simple consequence of the network's degree distribution. Panel B shows the results for a pollination mutualistic network (see M\_PL\_001 in Table \ref{tab:netdescr} of the Appendix \ref{app2}). The system exhibits a clear IBN structure that cannot be detected through the maximization of modularity. From the results in panels A and B, it is worth remarking that the observation of IBN structures in ecosystems with different types of interactions demands a reconsideration of which patterns should or should not be expected in them. Panel C shows the results for a urban user-service network (see Chennai in Table \ref{tab:netdescr} of the Appendix \ref{app2}). We observe again that global nestedness fails to characterize the predominant organization of the system, i.e.~an IBN structure.
Panel D shows the results for a unipartite network representing friendship relations in a Dutch school class (see c2 in Table \ref{tab:netdescr} of the Appendix \ref{app2}). The conclusions of the analysis of this network are similar to the ones in Panel A. 

\begin{figure*}[t]
	\begin{center}
		\includegraphics[width=1.5\columnwidth]{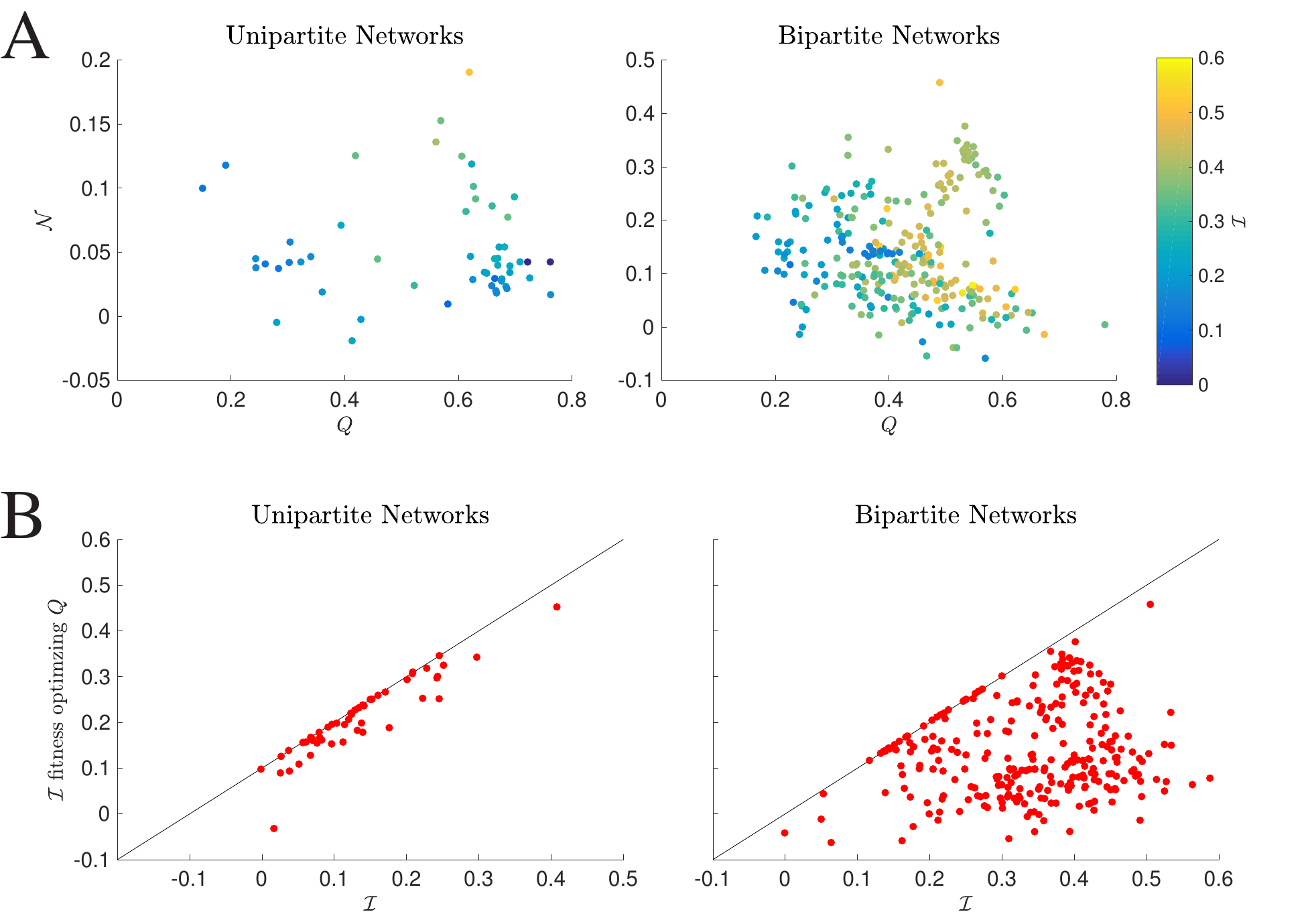}
	\end{center}
	\caption{(A) Scatter plot confronting modularity and nestedness with the $\mathcal{I}$ measure (color coded). Each point represents either an ecological, urban or social network (left panel: unipartite networks; right panel: bipartite networks). Note that bipartite networks have been analyzed under the formulation of Barber's modularity~\cite{barber2007modularity}. (B) Comparison of in-block nestedness value obtained optimising modularity or directly $\mathcal{I}$, as we propose.}
	\label{fig:real_results}
\end{figure*}

While Fig.~\ref{fig:real_results2} is helpful to get an intuition of how $\mathcal{I}$, $Q$ and $\mathcal{N}$ work, these representations may be misleading. For example, the degree-ranked representation in Panel B (right) conveys the idea that this particular plant-pollinator network is clearly (and globally) nested --but the other two arrangements are qualitatively convincing as well. For this reason, we have systematically compared the results of $\mathcal{N}$ and $Q$, on one side, and $\mathcal{I}$, on the other, for the whole set of real networks mentioned above. In Fig.~\ref{fig:real_results}A, two color-coded scatter plots are shown for uni- (left) and bipartite (right) networks. 
Strikingly, modularity $Q$ and in-block nestedness $\mathcal{I}$ are not strongly correlated in real datasets. 
Networks that exhibit small or intermediate values of modularity (compatible with those of a random network~\cite{guimera2004modularity}) may show high $\mathcal{I}$, regardless of the $\mathcal{N}$ score. Also, large values of $Q$ --which unsurprisingly display nestedness $\approx 0$-- indeed exhibit both large and small $\mathcal{I}$ scores as well. Beyond the uncorrelated behavior between the three descriptors, what surfaces here is the fact that when analyzing data we may be overlooking a relevant pattern --IBN structure-- just because two partial views of it ($\mathcal{N}$ and $Q$ taken independently) appear to be non-significant. 

Further, in Fig.~\ref{fig:real_results}B we show the value of $\mathcal{I}$ for partitions obtained by maximizing $Q$ confronted to the maximization of $\mathcal{I}$ itself. This plot evidences that modularity optimization may sometimes render partitions which do have some in-block organization (near the diagonal), but most often it is blind to it.
This result highlights that using an approach where modularity is maximized, to successively evaluate nestedness  within the blocks identified (i.e. the approach in \cite{lewinsohn2006structure,flores2013multi}), is not able to unveil the IBN structure in most real-world networks.  

\section{Conclusions}
\label{sec:conclusion}
The emergence of structural patterns in complex networks is a consequence of the dynamics that take place on them. While the ultimate goal is to understand how these dynamics operate, this is not feasible until the correct methods to identify those patterns are available. The increasing evidence that nestedness and modularity appear in many empirical contexts; the already abundant hints that they may appear together; and the importance of both to disentangle how they affect --and are affected by-- the evolution of a system, prompt the need of rethinking the strategy to detect the occurrence of in-block nestedness.

Inspired by the NODF and modularity optimization, in this paper we have developed a methodology to detect in-block nested structures. The objective function at the core of this method naturally embeds a suitable null model to discount the in-block nestedness of the network that can be ascribed to randomness.
Beyond the formal correctness of our formulation, demonstrated by means of a suitable benchmark to generate synthetic networks, we have shown that it overcomes the inherent limitations of nestedness and modularity (as independent methods) for this task. In structural terms, our approach can be interpreted as a generalization of the concept of nestedness (as expressed in NODF), re-framing it to the mesoscopic network scale. 
Along this line, the door remains open to further development. For example, slight modifications on our formulation generalize the notion of multiple core-periphery structures, which has been recently addressed from a different starting point~\cite{kojaku2017finding}. Other directions may be related to identification techniques (e.g.~stochastic block models ~\cite{peixoto2014hierarchical}) or the design of distinct, specialized null models ~\cite{jonhson2013factors,dominguez2015ranking}.

The analysis of real data has shown that many networks display in-block nested structure, regardless of their NODF and modularity scores. This finding suggests that previous works may have overlooked important features when discussing the organization of real systems. The existence of in-block nested structures affects the debate around population dynamics, in ecology specially, in terms of which patterns maximize survival~\cite{bastolla2009architecture}, and why. Our methodological contribution thus uncovers the need for models -- beyond host-parasite~\cite{beckett2013coevolutionary,leung2016conflicting} -- that explain how networks transition between possible configurations: from modular, to combined, to purely nested architectures, as suggested by the emergence of collective attention processes~\cite{borge2017emergence}; or from nested to combined architectures, as one would expect in a growing, but highly structured, system with increasing specialization. Noteworthy, it is not clear whether these processes are reversible, as they may respond to different system-wide adaptive processes. 

\section*{Acknowledgments}
CJT and MSM acknowledges the University of Z\"urich for support through the URPP Social Networks.
MSM also acknowledges financial support from the Swiss National Science Foundation Grant No. 200020-156188. 

\appendix
\section{Average degree of the synthetic network generation model} \label{sec:syntavdeg}

The synthetic in-block nested network generator we propose in the main text include two parameters to generate noise. Only as a summary, parameter $p$ controls the intra-compartment noise and parameter $\mu$ controls the inter-community noise. The noisy versions of the initial germinal noiseless in-block nested structure has the same average degree. We first show that adding intra-community noise does not alter the average number of edges. Under this situation $\mu = 0$ and Eq.~7 becomes

\begin{equation}
\begin{split}
	P(A^c_{ij}) &=((1-p) + pp_{r})\Theta(jN_c - f_n(iN_c)) +\\
    &+  p _r(1-\Theta(jN_c - f_n(iN_c)))
\end{split}
\end{equation}
and the expected number of edges within a compartment is $\langle E_c \rangle = \sum\limits_{ij}^{N_c} P(A^c_{ij})$. Reorganizing common terms of the sums we obtain that

\begin{equation}\label{averageDegreeCompartment}
\begin{split}
	\langle E_c \rangle &= \sum\limits_{ij}^{N_c} P(A_{ij}) =\\
    &=((1-p) + pp_{r})\sum\limits_{ij}^{N_c}\Theta(jN_c - f_n(iN_c)) \\
    &+  p _r\sum\limits_{ij}^{N_c}(1-\Theta(jN_c - f_n(iN_c)))
\end{split}
\end{equation}
The heaviside step function is one when a link was existing in the original noiseless compartment. Thus, by definition, $E_c=\sum\limits_{ij}^{N_c}\Theta(jN_c - f_n(iN_c))$, and consequently $N_c-E_c=\sum\limits_{ij}^{N_c}(1-\Theta(jN_c - f_n(iN_c)))$. Substituting, this to Eq. \ref{averageDegreeCompartment} leads to

\begin{equation}\label{block_num_edges}
\begin{split}
	\langle E_c \rangle &= (1-p + pp_{r})E_c +  p _r(N_c^2-E_c)=\\
    &= E_c - pE + p_r(N_c-E_c+pE_c)=E_c
\end{split}
\end{equation}
We now consider the case where $\mu\neq0$. The expected amount of edges within the full in-block nested structure is
\begin{equation}\label{}
\begin{split}
	\langle E \rangle &= B \sum\limits_{ij}^{N_c} P(A_{ij}^c) +\\
    &+ B(B-1)\sum\limits_{ij}^{N_c} P(A_{ij}^o) =\\
	&= B \(1-p_i\) \sum\limits_{ij}^{N_c}\[((1-p) + pp_{r})\Theta(jN_c - f_n(iN_c))\] +\\
    &+B \(1-p_i\) \sum\limits_{ij}^{N_c}\[p_r\Theta(f_n(iN_c) - jN_c)\] \\
	&+B(B-1)\sum\limits_{ij}^{N_c} \frac{E_c\mu}{N_c^2B} =\\
	&= B \(1-p_i\) E_c +B(B-1)\frac{\mu E_c}{N_c^2B} \sum\limits_{ij}^{N_c} 1 =\\
	&= B \(1-\frac{\mu(B-1)}{B}\) E_c + (B-1)\mu E_c = BE_c
\end{split}
\end{equation}

{\hfill \textit{\textbf{Q.E.D.}}}

\section{Real datasets used in the experiments}
\label{app2}
The set of real networks used throughout the article comprise ecological and social systems. The largest subset --ecological networks \cite{weboflife}-- represent mutualistic and competitive systems, including macroscopic and microscopic environments. The analyzed social systems include social communication networks such as face-to-face interactions, e-mail contacts and Twitter messages; urban systems such as user check-ins to city services (museum, market, restaurant, etc.); technological systems such as cooperative software development projects, where we account which files each user works on. Some of these networks have been previously shown to exhibit nestedness and modularity jointly \cite{olesen2007modularity,flores2013multi,borge2017emergence}. However, none of them have not been previously analyzed in the proposed setting.
\newpage
    \begin{longtable*}{| l | c | c | c | c | c |}
    \hline
    Name & M & N & Connectance & Type & Relation type \\ \hline
    \hline
    A\_HP\_002 & 24 & 18 & 0.22 & bipartite & Ecological Host-Parasite\\ \hline 
    A\_HP\_003 & 9 & 23 & 0.52 & bipartite & Ecological Host-Parasite\\ \hline 
    A\_HP\_006 & 37 & 16 & 0.21 & bipartite & Ecological Host-Parasite\\ \hline 
    A\_HP\_008 & 24 & 8 & 0.19 & bipartite & Ecological Host-Parasite\\ \hline 
    A\_HP\_009 & 22 & 14 & 0.31 & bipartite & Ecological Host-Parasite\\ \hline 
    A\_HP\_010 & 31 & 18 & 0.16 & bipartite & Ecological Host-Parasite\\ \hline 
    A\_HP\_013 & 22 & 11 & 0.43 & bipartite & Ecological Host-Parasite\\ \hline 
    A\_HP\_018 & 21 & 15 & 0.40 & bipartite & Ecological Host-Parasite\\ \hline 
    A\_HP\_020 & 17 & 16 & 0.39 & bipartite & Ecological Host-Parasite\\ \hline 
    A\_HP\_022 & 18 & 16 & 0.24 & bipartite & Ecological Host-Parasite\\ \hline 
    A\_HP\_025 & 40 & 18 & 0.15 & bipartite & Ecological Host-Parasite\\ \hline 
    A\_HP\_026 & 18 & 15 & 0.53 & bipartite & Ecological Host-Parasite\\ \hline 
    A\_HP\_027 & 30 & 17 & 0.21 & bipartite & Ecological Host-Parasite\\ \hline 
    A\_HP\_029 & 34 & 15 & 0.15 & bipartite & Ecological Host-Parasite\\ \hline 
    A\_HP\_031 & 31 & 25 & 0.22 & bipartite & Ecological Host-Parasite\\ \hline 
    A\_HP\_033 & 25 & 22 & 0.36 & bipartite & Ecological Host-Parasite\\ \hline 
    A\_HP\_037 & 21 & 17 & 0.25 & bipartite & Ecological Host-Parasite\\ \hline 
    A\_HP\_042 & 32 & 21 & 0.12 & bipartite & Ecological Host-Parasite\\ \hline 
    A\_HP\_043 & 29 & 9 & 0.28 & bipartite & Ecological Host-Parasite\\ \hline 
    A\_HP\_044 & 26 & 27 & 0.28 & bipartite & Ecological Host-Parasite\\ \hline 
    A\_HP\_046 & 39 & 17 & 0.30 & bipartite & Ecological Host-Parasite\\ \hline 
    A\_HP\_047 & 26 & 11 & 0.35 & bipartite & Ecological Host-Parasite\\ \hline 
    A\_HP\_050 & 35 & 27 & 0.24 & bipartite & Ecological Host-Parasite\\ \hline 
    A\_HP\_051 & 26 & 13 & 0.32 & bipartite & Ecological Host-Parasite\\ \hline 
    A\_PH\_004 & 22 & 52 & 0.16 & bipartite & Ecological Plant-Herbivore\\ \hline 
    A\_PH\_005 & 24 & 54 & 0.13 & bipartite & Ecological Plant-Herbivore\\ \hline 
    A\_PH\_006 & 88 & 6 & 0.22 & bipartite & Ecological Plant-Herbivore\\ \hline 
    A\_PH\_007 & 64 & 5 & 0.30 & bipartite & Ecological Plant-Herbivore\\ \hline 
    M\_PA\_003 & 15 & 24 & 0.12 & bipartite & Ecological Plant-Ant \\ \hline 
    M\_PA\_004 & 48 & 41 & 0.14 & bipartite & Ecological Plant-Ant\\ \hline 
    M\_PL\_001 & 101 & 84 & 0.04 & bipartite & Ecological Pollination\\ \hline 
    M\_PL\_002 & 64 & 43 & 0.07 & bipartite & Ecological Pollination\\ \hline 
    M\_PL\_003 & 25 & 36 & 0.09 & bipartite & Ecological Pollination\\ \hline 
    M\_PL\_004 & 102 & 12 & 0.14 & bipartite & Ecological Pollination\\ \hline 
    M\_PL\_005 & 275 & 96 & 0.03 & bipartite & Ecological Pollination\\ \hline 
    M\_PL\_006 & 61 & 17 & 0.14 & bipartite & Ecological Pollination\\ \hline 
    M\_PL\_007 & 36 & 16 & 0.15 & bipartite & Ecological Pollination\\ \hline 
    M\_PL\_008 & 38 & 11 & 0.25 & bipartite & Ecological Pollination\\ \hline 
    M\_PL\_009 & 118 & 24 & 0.09 & bipartite & Ecological Pollination\\ \hline 
    M\_PL\_010 & 76 & 31 & 0.19 & bipartite & Ecological Pollination\\ \hline 
    M\_PL\_012 & 55 & 29 & 0.09 & bipartite & Ecological Pollination\\ \hline 
    M\_PL\_013 & 56 & 9 & 0.20 & bipartite & Ecological Pollination \\ \hline 
    M\_PL\_014 & 81 & 29 & 0.08 & bipartite & Ecological Pollination \\ \hline 
    M\_PL\_015 & 666 & 131 & 0.03 & bipartite & Ecological Pollination \\ \hline 
    M\_PL\_016 & 179 & 26 & 0.09 & bipartite & Ecological Pollination \\ \hline 
    M\_PL\_017 & 79 & 25 & 0.15 & bipartite & Ecological Pollination \\ \hline 
    M\_PL\_018 & 105 & 39 & 0.09 & bipartite & Ecological Pollination \\ \hline 
    M\_PL\_019 & 85 & 40 & 0.08 & bipartite & Ecological Pollination \\ \hline 
    M\_PL\_020 & 91 & 20 & 0.10 & bipartite & Ecological Pollination \\ \hline 
    M\_PL\_021 & 677 & 91 & 0.02 & bipartite & Ecological Pollination \\ \hline 
    M\_PL\_022 & 45 & 21 & 0.09 & bipartite & Ecological Pollination \\ \hline 
    M\_PL\_023 & 72 & 23 & 0.08 & bipartite & Ecological Pollination \\ \hline 
    M\_PL\_025 & 44 & 13 & 0.25 & bipartite & Ecological Pollination \\ \hline 
    M\_PL\_026 & 54 & 105 & 0.04 & bipartite & Ecological Pollination \\ \hline 
    M\_PL\_027 & 60 & 18 & 0.11 & bipartite & Ecological Pollination \\ \hline 
    M\_PL\_028 & 139 & 41 & 0.07 & bipartite & Ecological Pollination \\ \hline 
    M\_PL\_029 & 118 & 49 & 0.06 & bipartite & Ecological Pollination \\ \hline 
    M\_PL\_030 & 53 & 28 & 0.07 & bipartite & Ecological Pollination \\ \hline 
    M\_PL\_031 & 49 & 48 & 0.07 & bipartite & Ecological Pollination \\ \hline 
    M\_PL\_032 & 33 & 7 & 0.28 & bipartite & Ecological Pollination \\ \hline 
    M\_PL\_033 & 34 & 13 & 0.32 & bipartite & Ecological Pollination \\ \hline 
    M\_PL\_034 & 128 & 26 & 0.09 & bipartite & Ecological Pollination \\ \hline 
    M\_PL\_035 & 36 & 61 & 0.08 & bipartite & Ecological Pollination \\ \hline 
    M\_PL\_037 & 40 & 10 & 0.18 & bipartite & Ecological Pollination \\ \hline 
    M\_PL\_038 & 42 & 8 & 0.24 & bipartite & Ecological Pollination \\ \hline 
    M\_PL\_039 & 51 & 17 & 0.15 & bipartite & Ecological Pollination \\ \hline 
    M\_PL\_040 & 43 & 29 & 0.09 & bipartite & Ecological Pollination \\ \hline 
    M\_PL\_041 & 43 & 31 & 0.11 & bipartite & Ecological Pollination \\ \hline 
    M\_PL\_043 & 82 & 28 & 0.11 & bipartite & Ecological Pollination \\ \hline 
    M\_PL\_044 & 609 & 110 & 0.02 & bipartite & Ecological Pollination \\ \hline 
    M\_PL\_045 & 26 & 17 & 0.14 & bipartite & Ecological Pollination \\ \hline 
    M\_PL\_046 & 44 & 16 & 0.39 & bipartite & Ecological Pollination \\ \hline 
    M\_PL\_047 & 186 & 19 & 0.12 & bipartite & Ecological Pollination \\ \hline 
    M\_PL\_048 & 236 & 30 & 0.09 & bipartite & Ecological Pollination \\ \hline 
    M\_PL\_049 & 225 & 37 & 0.07 & bipartite & Ecological Pollination \\ \hline 
    M\_PL\_050 & 35 & 14 & 0.18 & bipartite & Ecological Pollination \\ \hline 
    M\_PL\_051 & 90 & 14 & 0.13 & bipartite & Ecological Pollination \\ \hline 
    M\_PL\_052 & 39 & 15 & 0.16 & bipartite & Ecological Pollination \\ \hline 
    M\_PL\_053 & 294 & 99 & 0.02 & bipartite & Ecological Pollination \\ \hline 
    M\_PL\_054 & 318 & 113 & 0.02 & bipartite & Ecological Pollination \\ \hline 
    M\_PL\_055 & 195 & 64 & 0.03 & bipartite & Ecological Pollination \\ \hline 
    M\_PL\_056 & 365 & 91 & 0.03 & bipartite & Ecological Pollination \\ \hline 
    M\_PL\_057 & 883 & 114 & 0.02 & bipartite & Ecological Pollination \\ \hline 
    M\_PL\_058 & 81 & 32 & 0.12 & bipartite & Ecological Pollination \\ \hline 
    M\_PL\_060\_01 & 39 & 11 & 0.22 & bipartite & Ecological Pollination \\ \hline 
    M\_PL\_060\_02 & 38 & 12 & 0.23 & bipartite & Ecological Pollination \\ \hline 
    M\_PL\_060\_03 & 45 & 13 & 0.22 & bipartite & Ecological Pollination \\ \hline 
    M\_PL\_060\_04 & 46 & 21 & 0.14 & bipartite & Ecological Pollination \\ \hline 
    M\_PL\_060\_05 & 54 & 33 & 0.08 & bipartite & Ecological Pollination \\ \hline 
    M\_PL\_060\_06 & 45 & 26 & 0.08 & bipartite & Ecological Pollination \\ \hline 
    M\_PL\_060\_07 & 39 & 29 & 0.10 & bipartite & Ecological Pollination \\ \hline 
    M\_PL\_060\_08 & 28 & 19 & 0.14 & bipartite & Ecological Pollination \\ \hline 
    M\_PL\_060\_09 & 40 & 18 & 0.11 & bipartite & Ecological Pollination \\ \hline 
    M\_PL\_060\_10 & 25 & 14 & 0.15 & bipartite & Ecological Pollination \\ \hline 
    M\_PL\_060\_11 & 20 & 14 & 0.15 & bipartite & Ecological Pollination \\ \hline 
    M\_PL\_060\_12 & 26 & 11 & 0.21 & bipartite & Ecological Pollination \\ \hline 
    M\_PL\_060\_13 & 31 & 7 & 0.22 & bipartite & Ecological Pollination \\ \hline 
    M\_PL\_060\_14 & 37 & 11 & 0.24 & bipartite & Ecological Pollination \\ \hline 
    M\_PL\_060\_15 & 37 & 14 & 0.25 & bipartite & Ecological Pollination \\ \hline 
    M\_PL\_060\_16 & 39 & 17 & 0.17 & bipartite & Ecological Pollination \\ \hline 
    M\_PL\_060\_17 & 35 & 17 & 0.17 & bipartite & Ecological Pollination \\ \hline 
    M\_PL\_060\_18 & 28 & 20 & 0.13 & bipartite & Ecological Pollination \\ \hline 
    M\_PL\_060\_19 & 13 & 18 & 0.18 & bipartite & Ecological Pollination \\ \hline 
    M\_PL\_060\_22 & 31 & 13 & 0.16 & bipartite & Ecological Pollination \\ \hline 
    M\_PL\_060\_23 & 27 & 12 & 0.18 & bipartite & Ecological Pollination \\ \hline 
    M\_PL\_060\_24 & 24 & 14 & 0.14 & bipartite & Ecological Pollination \\ \hline 
    M\_PL\_061\_05 & 22 & 12 & 0.19 & bipartite & Ecological Pollination \\ \hline 
    M\_PL\_061\_06 & 24 & 11 & 0.22 & bipartite & Ecological Pollination \\ \hline 
    M\_PL\_061\_07 & 23 & 9 & 0.30 & bipartite & Ecological Pollination \\ \hline 
    M\_PL\_061\_19 & 21 & 11 & 0.19 & bipartite & Ecological Pollination \\ \hline 
    M\_PL\_061\_23 & 26 & 10 & 0.17 & bipartite & Ecological Pollination \\ \hline 
    M\_PL\_061\_38 & 23 & 10 & 0.21 & bipartite & Ecological Pollination \\ \hline 
    M\_PL\_061\_40 & 26 & 9 & 0.25 & bipartite & Ecological Pollination \\ \hline 
    M\_PL\_061\_45 & 23 & 11 & 0.20 & bipartite & Ecological Pollination \\ \hline 
    M\_PL\_061\_46 & 21 & 13 & 0.16 & bipartite & Ecological Pollination \\ \hline 
    M\_PL\_061\_47 & 26 & 9 & 0.24 & bipartite & Ecological Pollination \\ \hline 
    M\_PL\_063 & 9 & 55 & 0.25 & bipartite & Ecological Pollination \\ \hline 
    M\_SD\_002 & 9 & 31 & 0.43 & bipartite & Ecological Seed Dispersal\\ \hline 
    M\_SD\_003 & 16 & 25 & 0.17 & bipartite & Ecological Seed Dispersal\\ \hline 
    M\_SD\_004 & 20 & 34 & 0.14 & bipartite & Ecological Seed Dispersal\\ \hline 
    M\_SD\_005 & 13 & 25 & 0.15 & bipartite & Ecological Seed Dispersal\\ \hline 
    M\_SD\_006 & 15 & 21 & 0.16 & bipartite & Ecological Seed Dispersal\\ \hline 
    M\_SD\_007 & 7 & 72 & 0.28 & bipartite & Ecological Seed Dispersal\\ \hline 
    M\_SD\_010 & 14 & 50 & 0.33 & bipartite & Ecological Seed Dispersal\\ \hline 
    M\_SD\_012 & 29 & 35 & 0.14 & bipartite & Ecological Seed Dispersal\\ \hline 
    M\_SD\_013 & 19 & 36 & 0.29 & bipartite & Ecological Seed Dispersal\\ \hline 
    M\_SD\_014 & 17 & 16 & 0.44 & bipartite & Ecological Seed Dispersal\\ \hline 
    M\_SD\_015 & 27 & 5 & 0.64 & bipartite & Ecological Seed Dispersal\\ \hline 
    M\_SD\_016 & 61 & 24 & 0.34 & bipartite & Ecological Seed Dispersal\\ \hline 
    M\_SD\_018 & 32 & 29 & 0.07 & bipartite & Ecological Seed Dispersal\\ \hline 
    M\_SD\_019 & 40 & 169 & 0.10 & bipartite & Ecological Seed Dispersal\\ \hline 
    M\_SD\_020 & 33 & 25 & 0.18 & bipartite & Ecological Seed Dispersal\\ \hline 
    M\_SD\_021 & 28 & 18 & 0.26 & bipartite & Ecological Seed Dispersal\\ \hline 
    M\_SD\_022 & 110 & 207 & 0.05 & bipartite & Ecological Seed Dispersal\\ \hline 
    Women Event Participation& 18 & 14 & 0.35 & bipartite & Social \cite{davis2009deep} \\ \hline 
    Athens & 545 & 224 & 0.02 & bipartite & Foursquare Urban Checkins \cite{sarwat2014lars,levandoski2012lars} \\ \hline 
    Chennai & 280 & 279 & 0.04 & bipartite & Foursquare Urban Checkins \cite{sarwat2014lars,levandoski2012lars} \\ \hline 
    Mcp2000 & 146 & 1131 & 0.13 & bipartite & Github Project \\ \hline 
    animatecss & 77 & 254 & 0.03 & bipartite & Github Project\\ \hline 
    html5-boilerplate & 240 & 382 & 0.01 & bipartite & Github Project\\ \hline 
    javascript & 285 & 70 & 0.03 & bipartite & Github Project \\ \hline 
    UsrHstg\_0\_1439 & 49 & 48 & 0.08 & bipartite & Twitter User-Hashtag \cite{borge2017emergence}\\ \hline 
    UsrHstg\_10080\_11519 & 174 & 373 & 0.01 & bipartite & Twitter User-Hashtag \cite{borge2017emergence} \\ \hline 
    UsrHstg\_10800\_12239 & 160 & 317 & 0.01 & bipartite & Twitter User-Hashtag \cite{borge2017emergence} \\ \hline 
    UsrHstg\_11520\_12959 & 177 & 349 & 0.01 & bipartite & Twitter User-Hashtag \cite{borge2017emergence} \\ \hline 
    UsrHstg\_12240\_13679 & 158 & 332 & 0.01 & bipartite & Twitter User-Hashtag \cite{borge2017emergence} \\ \hline 
    UsrHstg\_12960\_14399 & 147 & 381 & 0.02 & bipartite & Twitter User-Hashtag \cite{borge2017emergence} \\ \hline 
    UsrHstg\_13680\_15119 & 170 & 404 & 0.01 & bipartite & Twitter User-Hashtag \cite{borge2017emergence} \\ \hline 
    UsrHstg\_14400\_15839 & 215 & 418 & 0.01 & bipartite & Twitter User-Hashtag \cite{borge2017emergence} \\ \hline 
    UsrHstg\_1440\_2879 & 148 & 344 & 0.01 & bipartite & Twitter User-Hashtag \cite{borge2017emergence} \\ \hline 
    UsrHstg\_15120\_16559 & 342 & 955 & 0.01 & bipartite & Twitter User-Hashtag \cite{borge2017emergence} \\ \hline 
    UsrHstg\_15840\_17279 & 378 & 1024 & 0.01 & bipartite & Twitter User-Hashtag \cite{borge2017emergence} \\ \hline 
    UsrHstg\_16560\_17999 & 283 & 872 & 0.01 & bipartite & Twitter User-Hashtag \cite{borge2017emergence} \\ \hline 
    UsrHstg\_17280\_18719 & 209 & 428 & 0.01 & bipartite & Twitter User-Hashtag \cite{borge2017emergence} \\ \hline 
    UsrHstg\_18000\_19439 & 192 & 378 & 0.01 & bipartite & Twitter User-Hashtag \cite{borge2017emergence} \\ \hline 
    UsrHstg\_18720\_20159 & 167 & 429 & 0.01 & bipartite & Twitter User-Hashtag \cite{borge2017emergence} \\ \hline 
    UsrHstg\_19440\_20879 & 253 & 709 & 0.01 & bipartite & Twitter User-Hashtag \cite{borge2017emergence} \\ \hline 
    UsrHstg\_20160\_21599 & 403 & 1024 & 0.01 & bipartite & Twitter User-Hashtag \cite{borge2017emergence} \\ \hline 
    UsrHstg\_20880\_22319 & 407 & 1024 & 0.01 & bipartite & Twitter User-Hashtag \cite{borge2017emergence} \\ \hline 
    UsrHstg\_21600\_23039 & 386 & 1024 & 0.01 & bipartite & Twitter User-Hashtag \cite{borge2017emergence} \\ \hline 
    UsrHstg\_2160\_3599 & 170 & 398 & 0.01 & bipartite & Twitter User-Hashtag \cite{borge2017emergence} \\ \hline 
    UsrHstg\_22320\_23759 & 441 & 1024 & 0.01 & bipartite & Twitter User-Hashtag \cite{borge2017emergence} \\ \hline 
    UsrHstg\_23040\_24479 & 418 & 1024 & 0.01 & bipartite & Twitter User-Hashtag \cite{borge2017emergence} \\ \hline 
    UsrHstg\_23760\_25199 & 354 & 1024 & 0.01 & bipartite & Twitter User-Hashtag \cite{borge2017emergence} \\ \hline 
    UsrHstg\_24480\_25919 & 314 & 844 & 0.01 & bipartite & Twitter User-Hashtag \cite{borge2017emergence} \\ \hline 
    UsrHstg\_25200\_26639 & 346 & 1024 & 0.01 & bipartite & Twitter User-Hashtag \cite{borge2017emergence} \\ \hline 
    UsrHstg\_25920\_27359 & 414 & 1024 & 0.01 & bipartite & Twitter User-Hashtag \cite{borge2017emergence} \\ \hline 
    UsrHstg\_26640\_28079 & 429 & 1024 & 0.01 & bipartite & Twitter User-Hashtag \cite{borge2017emergence} \\ \hline 
    UsrHstg\_27360\_28799 & 442 & 1024 & 0.01 & bipartite & Twitter User-Hashtag \cite{borge2017emergence} \\ \hline 
    UsrHstg\_28080\_29519 & 477 & 1024 & 0.01 & bipartite & Twitter User-Hashtag \cite{borge2017emergence} \\ \hline 
    UsrHstg\_28800\_30239 & 713 & 1024 & 0.01 & bipartite & Twitter User-Hashtag \cite{borge2017emergence} \\ \hline 
    UsrHstg\_2880\_4319 & 172 & 365 & 0.01 & bipartite & Twitter User-Hashtag \cite{borge2017emergence} \\ \hline 
    UsrHstg\_29520\_30959 & 909 & 1024 & 0.01 & bipartite & Twitter User-Hashtag \cite{borge2017emergence} \\ \hline 
    UsrHstg\_30240\_31679 & 870 & 1024 & 0.01 & bipartite & Twitter User-Hashtag \cite{borge2017emergence} \\ \hline 
    UsrHstg\_30960\_32399 & 960 & 1024 & 0.01 & bipartite & Twitter User-Hashtag \cite{borge2017emergence} \\ \hline 
    UsrHstg\_31680\_33119 & 1075 & 1024 & 0.01 & bipartite & Twitter User-Hashtag \cite{borge2017emergence} \\ \hline 
    UsrHstg\_32400\_33839 & 1115 & 1024 & 0.01 & bipartite & Twitter User-Hashtag \cite{borge2017emergence} \\ \hline 
    UsrHstg\_33120\_34559 & 1244 & 1024 & 0.01 & bipartite & Twitter User-Hashtag \cite{borge2017emergence} \\ \hline 
    UsrHstg\_33840\_35279 & 1420 & 1024 & 0.01 & bipartite & Twitter User-Hashtag \cite{borge2017emergence} \\ \hline 
    UsrHstg\_34560\_35999 & 1550 & 1024 & 0.01 & bipartite & Twitter User-Hashtag \cite{borge2017emergence} \\ \hline 
    UsrHstg\_35280\_36719 & 1480 & 1024 & 0.01 & bipartite & Twitter User-Hashtag \cite{borge2017emergence} \\ \hline 
    UsrHstg\_36000\_37439 & 1516 & 1024 & 0.01 & bipartite & Twitter User-Hashtag \cite{borge2017emergence} \\ \hline 
    UsrHstg\_37440\_38879 & 1535 & 1024 & 0.01 & bipartite & Twitter User-Hashtag \cite{borge2017emergence} \\ \hline 
    UsrHstg\_38160\_39599 & 1407 & 1024 & 0.01 & bipartite & Twitter User-Hashtag \cite{borge2017emergence} \\ \hline 
    UsrHstg\_38880\_40319 & 1347 & 1024 & 0.01 & bipartite & Twitter User-Hashtag \cite{borge2017emergence} \\ \hline 
    UsrHstg\_39600\_41039 & 1331 & 1024 & 0.01 & bipartite & Twitter User-Hashtag \cite{borge2017emergence} \\ \hline 
    UsrHstg\_40320\_41759 & 1180 & 1024 & 0.01 & bipartite & Twitter User-Hashtag \cite{borge2017emergence} \\ \hline 
    UsrHstg\_41040\_42479 & 1093 & 1024 & 0.01 & bipartite & Twitter User-Hashtag \cite{borge2017emergence} \\ \hline 
    UsrHstg\_41760\_43199 & 1059 & 1024 & 0.01 & bipartite & Twitter User-Hashtag \cite{borge2017emergence} \\ \hline 
    UsrHstg\_42480\_43919 & 1049 & 1024 & 0.01 & bipartite & Twitter User-Hashtag \cite{borge2017emergence} \\ \hline 
    UsrHstg\_43200\_44639 & 999 & 1024 & 0.01 & bipartite & Twitter User-Hashtag \cite{borge2017emergence} \\ \hline 
    UsrHstg\_4320\_5759 & 145 & 273 & 0.01 & bipartite & Twitter User-Hashtag \cite{borge2017emergence} \\ \hline 
    UsrHstg\_43920\_45359 & 899 & 1024 & 0.01 & bipartite & Twitter User-Hashtag \cite{borge2017emergence} \\ \hline 
    UsrHstg\_44640\_46079 & 820 & 1024 & 0.01 & bipartite & Twitter User-Hashtag \cite{borge2017emergence} \\ \hline 
    UsrHstg\_45360\_46799 & 611 & 1024 & 0.01 & bipartite & Twitter User-Hashtag \cite{borge2017emergence} \\ \hline 
    UsrHstg\_5040\_6479 & 149 & 278 & 0.01 & bipartite & Twitter User-Hashtag \cite{borge2017emergence} \\ \hline 
    UsrHstg\_5760\_7199 & 134 & 275 & 0.01 & bipartite & Twitter User-Hashtag \cite{borge2017emergence} \\ \hline 
    UsrHstg\_6480\_7919 & 116 & 231 & 0.02 & bipartite & Twitter User-Hashtag \cite{borge2017emergence} \\ \hline 
    UsrHstg\_7200\_8639 & 107 & 219 & 0.02 & bipartite & Twitter User-Hashtag \cite{borge2017emergence} \\ \hline 
    UsrHstg\_720\_2159 & 76 & 151 & 0.03 & bipartite & Twitter User-Hashtag \cite{borge2017emergence} \\ \hline 
    UsrHstg\_7920\_9359 & 107 & 237 & 0.02 & bipartite & Twitter User-Hashtag \cite{borge2017emergence} \\ \hline 
    UsrHstg\_8640\_10079 & 133 & 283 & 0.01 & bipartite & Twitter User-Hashtag \cite{borge2017emergence} \\ \hline 
    bunt0 & 31 & 31 & 0.01 & unipartite & Social contacts \cite{van1999friendship}\\ \hline 
    bunt1 & 32 & 32 & 0.14 & unipartite & Social contacts \cite{van1999friendship}\\ \hline 
    bunt3 & 32 & 32 & 0.21 & unipartite & Social contacts \cite{van1999friendship}\\ \hline 
    bunt4 & 32 & 32 & 0.25 & unipartite & Social contacts \cite{van1999friendship}\\ \hline 
    bunt5 & 32 & 32 & 0.29 & unipartite & Social contacts \cite{van1999friendship}\\ \hline 
    bunt6 & 32 & 32 & 0.24 & unipartite & Social contacts \cite{van1999friendship}\\ \hline 
    c1 & 26 & 26 & 0.19 & unipartite & Social contacts \cite{snijders2008introduction}\\ \hline 
    c2 & 26 & 26 & 0.25 & unipartite & Social contacts \cite{snijders2008introduction}\\ \hline 
    c3 & 26 & 26 & 0.31 & unipartite & Social contacts \cite{snijders2008introduction}\\ \hline 
    c4 & 26 & 26 & 0.28 & unipartite & Social contacts \cite{snijders2008introduction}\\ \hline 
    E-mail contacts (Milchaski) & 196 & 196 & 0.10 & unipartite & Social contacts\\ \hline 
    stu98t0 & 31 & 31 & 0.01 & unipartite & Social contacts \cite{van1999friendship}\\ \hline 
    stu98t2 & 34 & 34 & 0.27 & unipartite & Social contacts \cite{van1999friendship}\\ \hline
    stu98t3 & 34 & 34 & 0.36 & unipartite & Social contacts \cite{van1999friendship}\\ \hline 
    stu98t5 & 34 & 34 & 0.35 & unipartite & Social contacts \cite{van1999friendship}\\ \hline 
    stu98t6 & 34 & 34 & 0.35 & unipartite & Social contacts \cite{van1999friendship}\\ \hline 
    Primary School & 242 & 242 & 0.28 & unipartite & Social contacts \cite{stehle2011high} \\ \hline         
    High School 2011 & 126 & 126 & 0.22 & unipartite & Social contacts \cite{fournet2014contact} \\ \hline 
    High School 2012 & 180 & 180 & 0.14 & unipartite & Social contacts \cite{fournet2014contact} \\ \hline 
    E-mail contacts & 1133 & 1133 & 0.01 & unipartite & Social \cite{guimera2003self} \\ \hline 
    zachary & 34 & 34 & 0.13 & unipartite & Social \cite{zachary1977information} \\ \hline 
    Enron E-mail contacts M12 & 139 & 139 & 0.02 & unipartite & Social \cite{enron} \\ \hline 
    Enron E-mail contacts M13 & 287 & 287 & 0.01 & unipartite & Social \cite{enron} \\ \hline 
    Enron E-mail contacts M14 & 383 & 383 & 0.01 & unipartite & Social \cite{enron} \\ \hline 
    Enron E-mail contacts M15 & 420 & 420 & 0.01 & unipartite & Social \cite{enron} \\ \hline 
    Enron E-mail contacts M16 & 286 & 286 & 0.01 & unipartite & Social \cite{enron} \\ \hline 
    Enron E-mail contacts M17 & 367 & 367 & 0.01 & unipartite & Social \cite{enron} \\ \hline 
    Enron E-mail contacts M18 & 418 & 418 & 0.01 & unipartite & Social \cite{enron} \\ \hline 
    Enron E-mail contacts M19 & 1052 & 1052 & 0.00 & unipartite & Social \cite{enron} \\ \hline 
    Enron E-mail contacts M20 & 1285 & 1285 & 0.00 & unipartite & Social \cite{enron} \\ \hline 
    Enron E-mail contacts M21 & 2485 & 2485 & 0.00 & unipartite & Social \cite{enron} \\ \hline 
    Enron E-mail contacts M22 & 2477 & 2477 & 0.00 & unipartite & Social \cite{enron} \\ \hline 
    Enron E-mail contacts M23 & 2081 & 2081 & 0.00 & unipartite & Social \cite{enron} \\ \hline 
    Enron E-mail contacts M24 & 2158 & 2158 & 0.00 & unipartite & Social \cite{enron} \\ \hline 
    Enron E-mail contacts M25 & 3106 & 3106 & 0.00 & unipartite & Social \cite{enron} \\ \hline 
    Enron E-mail contacts M26 & 3479 & 3479 & 0.00 & unipartite & Social \cite{enron} \\ \hline 
    Enron E-mail contacts M27 & 3491 & 3491 & 0.00 & unipartite & Social \cite{enron} \\ \hline 
    Enron E-mail contacts M28 & 3990 & 3990 & 0.00 & unipartite & Social \cite{enron} \\ \hline 
    Enron E-mail contacts M29 & 4291 & 4291 & 0.00 & unipartite & Social \cite{enron} \\ \hline 
    Enron E-mail contacts M30 & 5138 & 5138 & 0.00 & unipartite & Social \cite{enron} \\ \hline 
    Enron E-mail contacts M31 & 4793 & 4793 & 0.00 & unipartite & Social \cite{enron} \\ \hline 
    Enron E-mail contacts M32 & 4081 & 4081 & 0.00 & unipartite & Social \cite{enron} \\ \hline 
    Enron E-mail contacts M33 & 3810 & 3810 & 0.00 & unipartite & Social \cite{enron} \\ \hline 
    Enron E-mail contacts M34 & 4341 & 4341 & 0.00 & unipartite & Social \cite{enron} \\ \hline 
    Enron E-mail contacts M37 & 7287 & 7287 & 0.00 & unipartite & Social \cite{enron} \\ \hline 
    Enron E-mail contacts M38 & 5013 & 5013 & 0.00 & unipartite & Social \cite{enron} \\ \hline 
    Enron E-mail contacts M39 & 4584 & 4584 & 0.00 & unipartite & Social \cite{enron} \\ \hline 
    Enron E-mail contacts M40 & 4702 & 4702 & 0.00 & unipartite & Social \cite{enron} \\ \hline 
    Enron E-mail contacts M43 & 6735 & 6735 & 0.00 & unipartite & Social \cite{enron} \\ \hline 
    Enron E-mail contacts M44 & 2974 & 2974 & 0.00 & unipartite & Social \cite{enron} \\ \hline 
    Enron E-mail contacts M45 & 3109 & 3109 & 0.00 & unipartite & Social \cite{enron} \\ \hline 
    Enron E-mail contacts M46 & 2231 & 2231 & 0.00 & unipartite & Social \cite{enron} \\ \hline 
    Enron E-mail contacts M47 & 1639 & 1639 & 0.00 & unipartite & Social \cite{enron} \\ \hline 
    Enron E-mail contacts M48 & 313 & 313 & 0.01 & unipartite & Social \cite{enron} \\ \hline 
    Enron E-mail contacts M49 & 1028 & 1028 & 0.00 & unipartite & Social \cite{enron} \\ \hline 
    Enron E-mail contacts M50 & 306 & 306 & 0.01 & unipartite & Social \cite{enron} \\ \hline 
    \caption{Details of the ecological and social networks used in the main document.}
    \label{tab:netdescr}
    \end{longtable*}


\begin{thebibliography}{51}
\bibitem [{bip()}]{bipartite}%
  \BibitemOpen
  \href@noop {} {}\bibinfo {note} {In the case of bipartite networks, the
  comparison must be done between nodes of the same class.}\BibitemShut {Stop}%
\end{thebibliography}
%

\end{document}